\documentclass[twocolumn,aps,prl,reprint,groupedaddress]{revtex4}
\usepackage{graphicx}
\usepackage{color}
\usepackage{amsmath}

\newcommand{\be}{\begin{equation}}
\newcommand{\ee}{\end{equation}}
\newcommand{\bea}{\begin{eqnarray}}
\newcommand{\eea}{\end{eqnarray}}

\newcommand{\p}{\partial}

\newcommand{\la}{\langle}
\newcommand{\ra}{\rangle}
\newcommand{\lb}{\left[}
\newcommand{\rb}{\right]}
\newcommand{\lp}{\left(}
\newcommand{\rp}{\right)}

\renewcommand{\vec}[1]{{\bf #1}}
\renewcommand{\hat}[1]{{\widehat #1}}

\begin{document}

\title{Electron Interactions and Gap Opening in Graphene Superlattices}
\author{Justin C. W. Song$^{1,2}$}
\author{Andrey V. Shytov$^3$}
\author{Leonid S. Levitov$^1$}
 \affiliation{$^1$ Department of Physics, Massachusetts Institute of Technology, Cambridge, Massachusetts 02139, USA}
 \affiliation{$^2$ School of Engineering and Applied Sciences, Harvard University, Cambridge, Massachusetts 02138, USA}
\affiliation{$^3$ School of Physics, University of Exeter, Stocker Road, Exeter EX4 4QL, U.K.}
\begin{abstract}
We develop a theory of interaction effects in graphene superlattices, where tunable superlattice periodicity can be used as a knob to control the gap at the Dirac point. Applied to graphene on hexa-boron-nitride (G/h-BN), our theory predicts substantial many-body enhancement of this gap. Tunable by the moir\'e superlattice periodicity, 
a few orders of magnitude enhancement is reachable under optimal conditions. The Dirac point gap enhancement can be much larger than that of the minigaps opened by Bragg scattering at principal superlattice harmonics.
 This naturally explains the conundrum of large Dirac point gaps recently observed in G/h-BN heterostructures and their tunability by the G/h-BN twist angle.
\end{abstract} 
\pacs{}
\maketitle

Opening up a bandgap in graphene promises to unlock a host of tantalizing new physics\cite{xiao2007,semenoff2008}. It will also enable its technological adoption\cite{avouris2007}. 
Recent attempts to open a gap at the Dirac point (DP)
can be broadly classed into two principal strategies: (a) spontaneous excitonic gap from electron-electron interactions \cite{pisarki1984, kotov2012}, and (b) inducing A-B sub-lattice asymmetry through an external potential (e.g. substrate) \cite{mattausch2007,varchon2007,pankratov,giovannetti2007,zhou2007, kindermann2012,wallbank}. However, unlike bilayer graphene where (a) succeeds, 
the vanishing density of states in monolayer graphene suppresses interaction effects. Furthermore, it is experimentally challenging to create a commensurate potential on the lattice scale without generating disorder.
So far, neither of these approaches alone have succeeded.

Here we propose a synergistic approach that relies on strength drawn from combining (a) and (b).
As we will see, fully in line with the 
adage ``the whole is greater than the sum of its parts," the interaction-enhanced Bragg scattering
by a relatively weak superlattice potential can lead to large gap values. We show that enhancements 
can be substantial 
under realistic conditions. Further, the enhancement is tunable by superlattice wavelength, opening the doorway to engineering interaction effects in graphene.

Here we apply these ideas to G/h-BN superlattices\cite{Park2008_NatPhys,Park2008_PRL,guinea2010,kindermann2012}.
Recent measurements found diverging resistance at DP in a G/h-BN system\cite{amet2012,hunt2013,cory-aps2013}. The insulating state at DP is observed despite the extreme cleanness of 
the system and when long-range disorder due to charge puddles is screened by gates. Curiously, the large gaps (of order $300 {\rm K}$) depended on twist angle\cite{hunt2013}. This was unanticipated by non-interacting models \cite{kindermann2012}. As we show, these observations are explained naturally by our  approach.

\begin{figure}[!h]
\includegraphics[scale=0.32]
{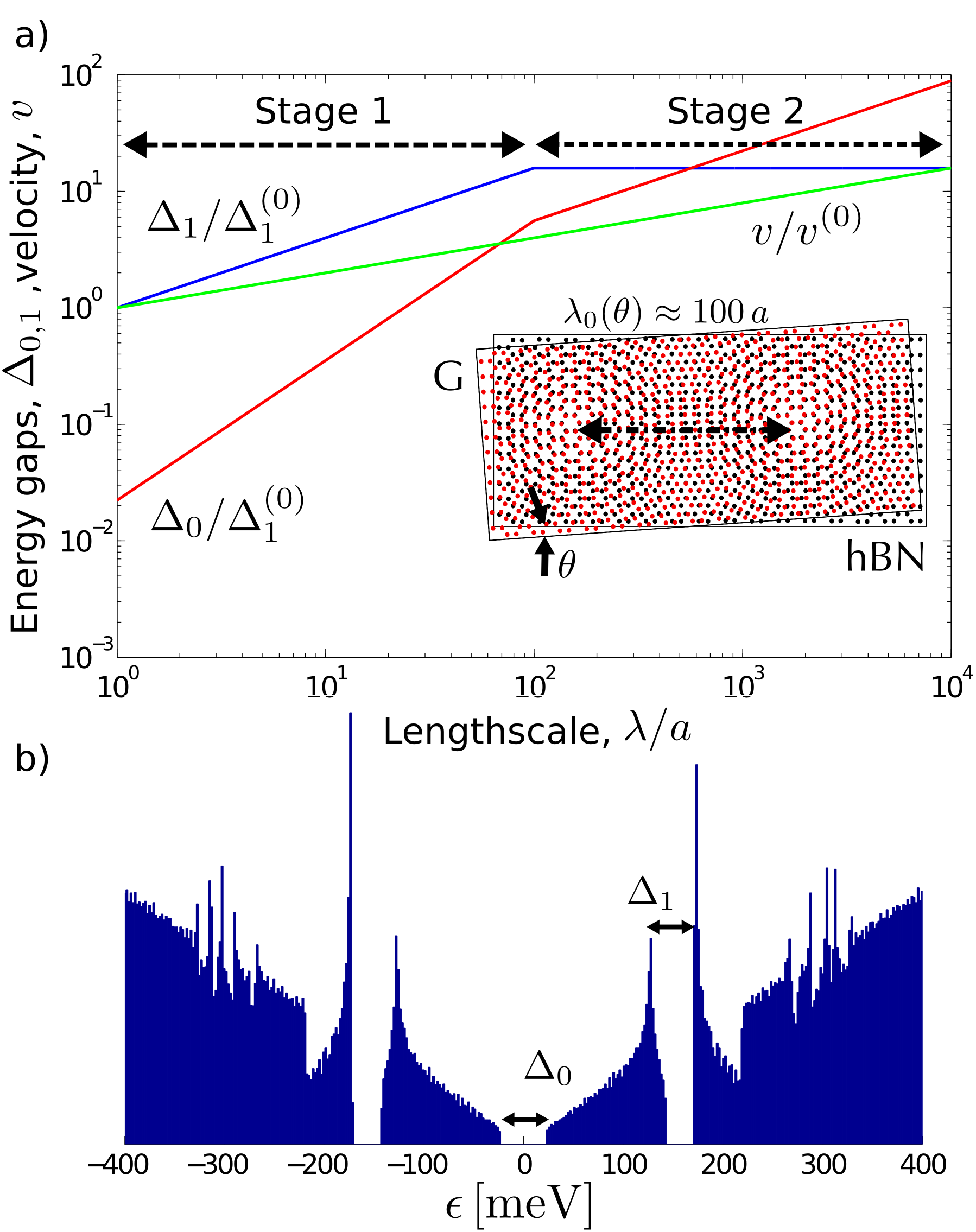}
\caption{(a) Interaction-induced enhancement of couplings to an incommensurate (moir\'e) superlattice potential. The superlattice spatial wavelength $\lambda$, tunable by the twist angle between G and h-BN (inset), controls the interaction effects via RG flow. Renormalization for the DP gap $\Delta_0$ and the gap $\Delta_1$ opened by Bragg scattering at the principal superlattice harmonics is shown by red and blue lines, respectively.
Renormalization proceeds in two stages. 
Both gaps are enhanced from the noninteracting values $\Delta_0^{(0)}$,  $\Delta_1^{(0)}$, however the gap $\Delta_0$ undergoes a much larger enhancement and, despite a small initial value, eventually overtakes $\Delta_1$. Both $\Delta_0$ and $\Delta_1$ grow faster than the carrier velocity $v$ (green line). 
b)  Density of states obtained using renormalized gaps and velocity. 
Parameters used: superlattice period $\lambda_0=14\,{\rm nm}$, scaling exponents $\beta=0.6$,  $\beta_v=0.3$, initial values $\Delta_0^{(0)}\approx 0.02\Delta_1^{(0)}$ (see text). 
}
\label{fig1}
\vspace{-6mm}
\end{figure}


The dependence on the twist angle arises because this angle controls the periodicity of G/h-BN moir\'e patterns (Fig.1 inset).
The wavelengths in moir\'e superlattices can be as long as 100 atomic distances,
depending on both the lattice mismatch 
and twist angle between graphene and h-BN, $\lambda_0(\theta)\approx\lambda_0\delta/(\theta^2+\delta^2)^{1/2}$ \cite{wallbank,kindermann2012}. 
In turn, the wavelength $\lambda$ controls 
the DP gap $\Delta_0$ enhancement via renormalization group (RG) flow (Fig.1).
As a result, $\Delta_0$ can be tuned by $\theta$. Our RG approach predicts a power-law scaling, 
\be
\label{eq:gap_angle}
\Delta_0(\theta) \propto \lp\lambda_0 (\theta)/a\rp^\gamma
,
\ee
giving a strong dependence on 
the twist angle, $\theta$. 
Here $a=1.42\,\AA$ is the carbon spacing and the value $\gamma$ is estimated below [see Eq.(\ref{eq:exponent})].
The twist angle dependent $\Delta_0$ is consistent with recent experimental observations \cite{hunt2013}.
Our estimates, based on a one-loop RG, show that DP gap values as large as $\Delta_0\approx 5$-$40{\rm meV}$ can be realized. The large enhancement predicted by RG originates from the fact that the DP gap arises at third order in a weak superlattice potential. While this suppresses the bare gap value, it also triples the scaling exponent $\beta$ describing the interaction-induced enhancement, see Eqs.(\ref{eq:beta_answer}),(\ref{eq:Delta0_stage1}).

The RG approach complements, in an important way, existing {\it ab initio} calculations for commensurate and lattice-matched graphene heterostructures. Recently, gaps at DP arising from the sublattice asymmetry of SiC\cite{mattausch2007,varchon2007,pankratov} and h-BN\cite{giovannetti2007} have been predicted; 
substrate induced gaps were observed in epitaxial graphene on SiC \cite{zhou2007}. Interaction-enhanced gaps 
were also analyzed in a commensurate structure \cite{kotov2009}. 
However, experimentally realistic graphene heterostructures are {\it incommensurate and lattice-mismatched} with relevant superlattice lengthscales that exceed the atomic scale by almost two orders of magnitude. 
Such large registrations are beyond the capability of {\it ab initio} techniques, which are typically limited to cell sizes of tens of atoms. 
On the other hand, the RG approach is ideally suited for describing long wavelength behavior.
In this work, we set out to understand the interplay of interactions and incommensurability. 
While we focus on G/hBN systems, our approach applies equally to other closely matched substrates such as SiC.

The origin of a strong effect of interactions on the band structure can be understood as follows.  Interacting Dirac particles respond 
 very differently 
 to a scalar external potential which is sublattice-blind than 
to a pseudospin-dependent (`colored') potential reflecting the $A$-$B$ sublattice modulation. In the first case, interactions generate a polarization that screens the potential. In the second case,  interactions generate sublattice correlations that amplify the potential. 
The interaction-enhanced colored 
potential leads to pseudospin-dependent Bragg scattering which generates a gap at DP, $\Delta_0$.
Further, because the Dirac mass term is a relevant perturbation in the RG 
sense, this gap undergoes a giant interaction-induced enhancement. Interestingly, 
the resulting $\Delta_0$ value can exceed the side gaps opened by Bragg scattering at the principal superlattice harmonics, see Fig.\ref{fig1}. 

Due to long-period spatial oscillations in the moir\'e  superlattice, the RG flow proceeds in two separate stages: stage 1 describing renormalization at lengthscales up to the moir\'e wavelength, $\lambda<\lambda_0$, and stage 2 describing lengthscales $\lambda>\lambda_0$. The lengthscale at which RG terminates is controlled by the screening length which is set by the distance to the gates when those are metallic, as in Ref.\cite{dean2010}, or by the screening length in the gate if the latter is realized by a proximal graphene layer, as in Ref.\cite{ponomarenko2011}. Stage 1 acts as a `booster,' generating an enhancement of  energy gaps which is much greater than that of carrier velocity. 
In stage 2 the growth of  the side gap $\Delta_1$ stalls whereas the gap $\Delta_0$ continues to grow. As illustrated in Fig.\ref{fig1}, the gap $\Delta_0$ 
can eventually overtake $\Delta_1$ even if the latter starts with a larger microscopic bare value. This unusual hierarchy of energy scales, $\Delta_0 > \Delta_1$, can serve as a telltale sign of interaction-assisted gap opening. 

Turning to the analysis, since the moir\'e superlattice period is much larger than the lattice constant, $\lambda_0\gg a$, states in valleys $K$ and $K'$ are effectively decoupled. Hence we can describe each valley by a continuum Hamiltonian
with a pseudospin dependence reflecting $A$-$B$ sublattice modulation, and an oscillatory position dependence reflecting the superlattice periodicity:
\bea
&&H_0=
\lp\begin{array}{cc}u_{11}f(\vec x) & vp_- +u_{12}f(\vec x)\\  vp_+ +u_{21}f(\vec x)& u_{22}f(\vec x)\end{array}\rp
,\quad
\\
&&
f(\vec x)=2\sum_{s=1,2,3}\cos(\vec b_s\vec x+\phi_s)
,
\eea
$p_\pm=p_x\pm ip_y$. 
Here $\vec b_s$ ($s=1,2,3$) are  three Bravais vectors of the triangular superlattice oriented at $60^{\rm o}$ angles relative to each other. 
Introducing pseudospin Pauli matrices $\sigma_{1,2,3}$ 
and adding long-range interactions we write
\bea \nonumber
{\cal H}=\!\!\!\!&& \int d^2x  \sum_{i=1}^N\psi_i^\dagger(\vec x) 
\lb v\boldsymbol\sigma\cdot \vec p+m_3(\vec x)\sigma_3+m_0(\vec x)\rb \psi_i(\vec x)
\\ \label{eq:H_total} && 
+\frac12\int d^2x\int d^2x' \frac{e^2}{\kappa|\vec x-\vec x'|} n(\vec x)n(\vec x')
,
\eea
where $m_3=\frac12(u_{11}-u_{22})f(\vec x)$, $m_0=\frac12(u_{11}+u_{22})f(\vec x)$. Here $N=4$ is the number of spin/valley flavors, 
and $n(\vec x)=\sum_{i=1...N}\psi_i^\dagger(\vec x)\psi_i(\vec x)$ is particle density. The off-diagonal terms $u_{12}$ and $u_{21}$, which describe a gauge-field coupling generated by strain, can be incorporated in the $v\sigma\vec p$ term. However, {\it ab initio} studies\cite{sachs2011} indicate that it is a small contribution compared to  $u_{11}$ and $u_{22}$.

The  amplitude of sublattice modulation can be inferred from {\it ab initio} calculations\cite{giovannetti2007} predicting $6(u_{11}-u_{22})\approx 
53\,{\rm meV}$ for the equal-period case, $\vec b=0$. This gives amplitudes $m_3$ of individual Bragg harmonics 
which are more than 20 times smaller than the kinetic energy at the Bragg vector, $\epsilon_0=\frac12\hbar v |\vec b|\approx 150\,{\rm meV}$ estimated for the largest superlattice period $\lambda_0=14\,{\rm nm}$ \cite{yankowitz2012}. Hence we can employ perturbation theory in the small ratio $m_3/\epsilon_0$. 

The RG analysis of this Hamiltonian predicts that the $m_3$ harmonics grow under RG, whereas the $m_0$ harmonics do not grow (see below). Therefore, even if the microscopic values $m_3^{(0)}$ and $m_0^{(0)}$ are comparable, the bandgaps will be dominated by the $m_3$ harmonics.  We can therefore estimate the bandgap opening at the edge of the superlattice Brillouin zone as $\Delta_1=2m_3$. 

\begin{figure}
\includegraphics[scale=0.3]{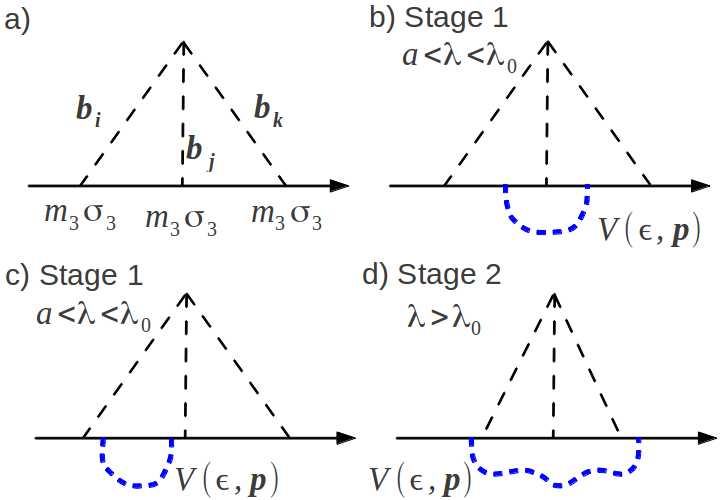}
\caption{(a) Self-energy describing gap opening at the Dirac point due to the bare sublattice-asymmetric superlattice potential [see Eq.(\ref{eq:Delta_0^0})].  (b,c,d) Log-divergent diagrams contributing to gap renormalization at one loop. Here solid black lines represent the electron Greens function $G(\epsilon, \vec p)$, dashed lines represent coupling to the sublattice-asymmetric potential [the term $m_3\sigma_3$ in Eq.(\ref{eq:H_total})], the wavy lines represent the Coulomb interaction.
Vertex renormalization  (b) and velocity renormalization (c) 
arise from integration over $|\vec b|\lesssim |\vec p|\lesssim p_0$ (stage 1), giving Eq.(\ref{eq:Delta0_stage1}). The contribution (d) arises from $|\vec p|<|\vec b|$ (stage 2), giving Eq.(\ref{eq:Delta0_stage2}).
}
\label{fig2}
\vspace{-5mm}
\end{figure}

Importantly, the interaction with the superlattice leads to gap opening at DP. 
While a single $m_3$ harmonic is sign-changing and 
cannot give rise to 
a gap by itself, $\la e^{i\vec b\vec x}\ra=0$, a combination of three different harmonics can open up a gap at DP. Choosing triplets of harmonics with the sum of their wave-vectors adding up
to zero, $\vec b_i +\vec b_j +\vec b_k = 0$, third-order perturbation theory in $m_3$ yields a constant sublattice-asymmetric term \footnote{In Eq.(\ref{eq:Delta_0^0}) we have assumed $\delta\phi=\phi_1+\phi_2+\phi_3=0$. For nonzero $\delta\phi$ we find the dependence $\Delta_0(\delta\phi)=\cos(\delta\phi)\Delta_0$. }
\be\label{eq:Delta_0^0}
H_0' =\!\!\!\!\sum_{\pm\vec b_i,\pm\vec b_k}\!\!\! m_3\sigma_3\frac1{v\boldsymbol{\sigma}\cdot\vec b_i}m_3\sigma_3\frac1{v\boldsymbol{\sigma}\cdot\vec b_k}m_3\sigma_3=-\frac12\Delta_0\sigma_3
,
\ee
where $\Delta_0=12m_3^3/(v|\vec b|)^2$ (the corresponding self-energy is shown in Fig.\ref{fig2}a).
Contributions similar to $H_0'$ can also arise when two out of three $m_3\sigma_3$ terms are replaced by $m_0$ harmonics. However, since these harmonics do not grow under RG (see below), these contributions are small. Since $m_3\ll \epsilon_0$, the predicted numerical value is small, $\Delta_0\ll\Delta_1$ (a similar observation was made in Ref.\cite{wallbank}). However, as we find shortly, the gap $\Delta_0$ undergoes a giant boost due to interaction effects, growing faster than $\Delta_1$. As a result, the physical values $\Delta_0$ and $\Delta_1$ become comparable.

We describe the effect of interactions on the terms $m_0$ and $m_3$ in Eq. (\ref{eq:H_total}) using the RG approach developed in Refs.\cite{gonzalez1994,vafek2007,son2007}.
There are two distinct flavors of RG, a weak-coupling approach and a large-$N$ approach\cite{kotov2012}. Weak-coupling RG, which  uses $e^2/\hbar v\ll 1$ as an expansion parameter, does not account for screening of the Coulomb interaction and features logarithmic enhancements. The large-$N$ RG, in contrast, describes strong coupling, fully accounts for screening and produces power law enhancements (see below). The two approaches differ quantitatively, yet they lead to qualitatively similar results. 
Here we present detailed results for the more realistic large-$N$ approach.

We first treat the $m_0$ and $m_3$ terms in Eq. (\ref{eq:H_total}) as spatially uniform, ignoring their $\vec x$  dependence.
The approach is valid over the range of lengthscales $a<\lambda\lesssim \lambda_0$. Larger lengthscales, $\lambda\gtrsim \lambda_0$, will be analyzed below. Renormalization is found by dressing the $m_0$ and $m_3$ vertices in Feynman diagrams with vertex corrections and analyzing the corresponding log-divergent contributions. 
The vertex correction for the $m_0$ term is cancelled by a corresponding contribution to the Greens function residue, owing to the Ward identity that follows from gauge invariance. This can be seen more explicitly by analyzing the self-energy 
\be
\Sigma(\epsilon,\vec p)=-\int \frac{d\epsilon'd^2p'}{(2\pi)^3}V(\epsilon',\vec p')G(\epsilon-\epsilon',\vec p-\vec p')
\ee
with $G(\epsilon,\vec p)=1/(i\epsilon-v\boldsymbol{\sigma}\cdot \vec p-m_3\sigma_3-m_0)$,
and $V$ representing the dynamically screened interaction, 
\be
V(\epsilon,\vec p)=\frac{V_0(\vec p)}{1-NV_0(\vec p)\Pi(\epsilon,\vec p)}
,\quad
V_0(\vec p)=\frac{2\pi e^2}{\kappa |\vec p|}
.
\ee
Renormalization of the $m_0$ coupling, at linear order in $m_0$, is described by the quantity $\p_{i\epsilon}\Sigma+\p_{m_0}\Sigma$ which vanishes due to the form of $G(\epsilon,\vec p)$. 

We note that the cancellation of log-divergent contributions due to the Ward identity does not apply to reducible 
diagrams. The latter generate a vertex correction giving an effective dielectric constant $\tilde\kappa =[1-NV_0(\vec p)\Pi(\epsilon,\vec p)]\kappa$, which describes intrinsic screening of $m_0$ vertex by inter-band and intra-band polarization. 

Since $m_3$ vertex is distinct from $m_0$ vertex, we expect that log-divergent contributions  do not cancel. As a result, electrons become `colored,' i.e. their coupling to the moir\'e superlattice potential is dominated by pseudospin-dependent interactions. Renormalization of $m_3$ was analyzed in Ref.\cite{kotov2009}
giving a scaling exponent $\beta=16/\pi^2N$ which is two times larger than the value $\beta_v=8/\pi^2N$ found for velocity renormalization in Ref.\cite{son2007}. This leads to RG flow equations for the $m_3$ coupling and velocity: 
\be
\frac{d m_3}{d\xi} = \beta m_3, \quad 
\frac{d v}{d\xi} = \beta_v v
,\quad
\frac{2\pi}{\lambda_0}<|\vec p|<p_0
,
\label{eq:RG}
\ee
where $\xi=\ln\frac{p_0}{|\vec p|}$ is the RG time parameter, with $p_0\sim 2\pi /a$ the UV cutoff.  Interestingly, the relation $\beta=2\beta_v$ also holds in the weak-coupling approach. The RG flow, Eq.(\ref{eq:RG}), predicts a power-law enhancement to the $m_3$ harmonic and velocity for $a<\lambda<\lambda_0$:
\be
m_3= \big(\lambda /a\big) ^{\beta}m_3^{(0)}, \quad v= \big(\lambda /a \big) ^{\beta_v}v^{(0)}
.
\label{eq:m3andv}
\ee
Similarly, renormalization of $\Delta_0\propto m_3^3/v^2$ is obtained by adding the contributions shown in Fig.\ref{fig2}(b,c):
\bea\nonumber
&&\frac{\p\Sigma}{i\p \epsilon}+\frac{3\delta\Sigma}{\delta(\sigma_3 m_3)} -\frac{2\delta\Sigma}{\delta(v\boldsymbol{\sigma}\vec p)} =
\int \frac{d\epsilon'd^2p'}{(2\pi)^3}
\lb\frac{V(\epsilon',\vec p')}{(i\epsilon'-v\boldsymbol{\sigma}\vec p')^2}\right.
\\ \label{eq:beta_answer}
&&
-\frac{3V(\epsilon',\vec p')}{(i\epsilon'-v\boldsymbol{\sigma}\vec p')(i\epsilon'+v\boldsymbol{\sigma}\vec p')}
\\\nonumber
&&
\left. -\frac{2V(\epsilon',\vec p'){\epsilon'}^2}{(i\epsilon'-v\boldsymbol{\sigma}\vec p')^2(i\epsilon'+v\boldsymbol{\sigma}\vec p')^2}
\rb
 \approx (3\beta-2\beta_v)\ln\frac{p_0}{|\vec p|} 
,
\eea
where we integrate over $|\vec p|\le p'\le p_0$. Taking $N$ to be large, we find the value $32/\pi^2 N$ which is two times larger than the exponent $\beta$ found in Ref.\cite{kotov2009} and four times larger than the exponent $\beta_v$. As a result,  $\Delta_0$ grows under RG far faster than $\Delta_1$:
\be\label{eq:Delta0_stage1}
\Delta_0 =   \frac{12 m_3^3}{v^2 |\vec b|^2} = \big(\lambda/ a \big)^{3 \beta - 2\beta_v}\Delta_0^{(0)} 
. 
\ee
A ten-fold increase in $\Delta_1$ translates into a hundred-fold increase in $\Delta_0$, see Fig.\ref{fig1}.
Thus, despite a handicap due to a small initial value, Eq.(\ref{eq:Delta_0^0}), the physical values for $\Delta_0$ and $\Delta_1$ are in the same ballpark.

For the sake of generality,  and acknowledging an approximate character of the scaling dimensions obtained from one-loop RG, we shall leave $\beta$ and $\beta_v$ unspecified in the analytic expressions. An attempt to  experimentally determine scaling exponents was made recently in Ref.\cite{elias2011}, where a systematic change of the period of quantum oscillations with carrier density was interpreted in terms of Fermi velocity renormalization, giving a value $\beta_v=0.5$-$0.55$, which is considerably larger than the one-loop RG result, $\beta_v=\frac8{\pi^2 N}\approx 0.2$. 
This discrepancy is not yet understood. 
Yet, even a modest change in $\beta_v$ may strongly impact the RG flow for the gap. Indeed, the enhancement factor $\Delta_0(\lambda_0)/\Delta_0^{(0)}$ is of order $50$ for $\beta_v = 0.2$, and grows exponentially as $\beta_v$ increases. Since the $\beta_v$  value is probably underestimated by one-loop RG,  we pick a conservative value $\beta_v=0.3$ for the plots in Fig.\ref{fig1}, giving  $\beta=2\beta_v=0.6$ (with the superlattice period $\lambda_0 = 14 \, {\rm nm}$). This suffices to illustrate the dramatic character of gap enhancement due to interactions.


We note that the large gap enhancement in Eq.(\ref{eq:Delta0_stage1}) stems predominantly from the combinatorial factor of 3 in the exponent $3\beta-2\beta_v$. This factor reflects the general structure of coupling to superlattice, which generates the bare DP gap at third order in the coupling $m_3$. We therefore expect that, while the exponent $\beta$ value may change due to two-loop RG corrections, the exponent in Eq.(\ref{eq:Delta0_stage1}) will remain large enough to generate substantial gap enhancement.

At lengthscales larger than $\lambda_0$,
renormalization is suppressed by spatial oscillations in the $m_3$ term. In this regime $m_3$ stops flowing and the gap $\Delta_1$ stays constant, as shown by the horizontal line in Fig.\ref{fig1} (stage 2). In contrast, because the effective Hamiltonian $H_0'$ is $\vec x$-independent, 
modes with $|\vec p| < \frac{2\pi}{\lambda_0}$ continue to provide an enhancement to the gap at DP, giving
\be\label{eq:Delta0_stage2}
\Delta_0 (\lambda >\lambda_0) = \lp \lambda/\lambda_0\rp^{\beta}\Delta_0(\lambda_0)
.
\ee
The RG flow of $\Delta_0$ at stage 2 is described by the scaling exponent $\beta$ which is smaller than the value  $3\beta-2\beta_v$ found for stage 1. Nevertheless, the growth of $\Delta_0$ at stage 2 is still faster than that of velocity. 
As illustrated in Fig.1,  $\Delta_0$ can grow by several orders of magnitude, reaching $\Delta_1$ for large enough $\lambda$. For the parameters chosen in Fig.\ref{fig1}, $\Delta_0$ (red line) overtakes $\Delta_1$ (blue line), eventually making the gap at DP the largest gap in the system.

Renormalization of all the parameters in the system, including velocity $v$ and the gap $\Delta_0$, terminates either at an effective screening length set by the proximal gate or at a lengthscale generated self-consistently through a gap opening at DP. In the first case, the effective screening length is determined by the distance to the gate or by the screening length for the gate, whichever is larger. In the second case, realized for systems with remote gates or when the screening length is very large, RG terminates at a self-consistently defined lengthscale, $\lambda_*$, controlled by the gap opened at DP, 
\be
\lambda_*=2\pi/ q_*
,\quad
\hbar q_*=\Delta_0(\lambda_*)/v(\lambda_*)
,
\ee
where the  $\lambda$ dependence is obtained from RG flow. A similar approach was employed in Refs.\cite{levitov2001,levitov2003} to estimate interaction-enhanced gaps opened at DP in chiral metallic carbon nanotubes. Under realistic conditions, as discussed above, we expect $\Delta_0\ll\epsilon_0=\frac12 \hbar v|\vec b|$. Hence, the selfconsistent lengthscale satisfies 
$\lambda_*\gg\lambda_0$. In this case, plugging 
the dependence from Eqs.(\ref{eq:m3andv}),(\ref{eq:Delta0_stage2}), we find $(\lambda_*/\lambda_0)^{1+\beta-\beta_v}=h v(\lambda_0)/[\lambda_0\Delta_0(\lambda_0)]$, giving
\be\label{eq:Delta_max}
\Delta_0(\lambda_*)=\lp \frac{h v(\lambda_0)}{\lambda_0\Delta_0(\lambda_0)}\rp^\frac{\beta}{1+\beta-\beta_v}\Delta_0(\lambda_0)
\propto
\lp\frac{\lambda_0}{a}\rp^\gamma
,
\ee
$
\gamma=3\beta-2\beta_v-\frac{\beta(1+3\beta-3\beta_v)}{1+\beta-\beta_v}
$.
The predicted power-law dependence $\Delta_0$ vs. moir\'e wavelength $\lambda_0$, Eq.(\ref{eq:Delta_max}), can be used to directly probe the effects of interactions. 
Setting $\beta=2\beta_v$ (see discussion above), the expression for the scaling exponent simplifies as:
\be
\gamma=4\beta_v-\frac{2\beta_v(1+3\beta_v)}{1+\beta_v}=\frac{\beta(2-\beta)}{2+\beta}
\label{eq:exponent}
\ee
For $\beta=\frac{16}{\pi^2 N} =0.4$ (a lower bound for the exponent) we find $\gamma \approx 0.27$. 
This produces a characteristic angle dependent gap [see Eq.(\ref{eq:gap_angle})]. 

The predicted scaling, Eq.(\ref{eq:gap_angle}), can be tested by comparing the gap values measured in G/hBN systems with different twist angles, similar to the method used in Ref.\cite{yankowitz2012} for replica Dirac peaks. Renormalization effects are maximized for structures with near-perfect crystal axes alignment, such as those in recent experiments \cite{batterfly_manchester,hunt2013,cory-aps2013}. 
Indeed, as illustrated in Fig.\ref{fig1}a, $\Delta_0$ can be enhanced by a factor of up to $10^3$. A conservative estimate of the superlattice harmonic,  $m_3^{(0)} = \frac1{12}(53\, {\rm meV})$\cite{giovannetti2007}, would yield a gap as large as $\Delta_0 \approx 5-40\, {\rm meV}$ [using Eq.(\ref{eq:Delta_max}) and $\beta = 2\beta_v = 0.4 - 0.6$]. This is close to recently observed gap values \cite{amet2012,cory-aps2013,hunt2013}. 

Besides the large gap values, interactions generate anomalous hierarchy of gap sizes ($\Delta_0>\Delta_1$), and  a strong dependence of the gap $\Delta_0$ on the twist angle, Eq.(\ref{eq:gap_angle}). These effects provide clear experimental signatures for the proposed scenario. 
The effective interaction strength can be tuned by adjusting superlattice wavelength, which is done by rotating the G layer relative to the BN layer. 
This opens the doorway for realizing and exploring tunable interaction effects in graphene superlattices.

We thank D. Golhaber-Gordon, A. Geim, P. Jarillo-Herrero, J. Williams and A. Young for useful discussions.


\begin{thebibliography}{99}
\vspace{-3mm}


\bibitem{xiao2007} D. Xiao, W. Yao, and Q. Niu, Phys. Rev. Lett., {\bf 99} 236809 (2007).

\bibitem{semenoff2008} G. W. Semenoff, V. Semenoff, and F. Zhou, Phys. Rev. Lett., {\bf 101} 087204 (2008).

\bibitem{avouris2007} P. Avouris, Z. Chen, V. Perebeinos, Nat. Nano., {\bf 2} 605 (2007).

\bibitem{pisarki1984} R. D. Pisarski, Phys. Rev. D, {\bf 29} 2423. (1984).




\bibitem{kotov2012} V. N. Kotov, B. Uchoa, V. M. Pereira, F. Guinea, and A. H. Castro
Neto, 
Rev. Mod. Phys. {\bf 84} 1067 (2012).


\bibitem{wallbank}
J. R. Wallbank, A. A. Patel, M. Mucha-Kruczynski, A. K. Geim, V. I. Fal'ko, 
 Phys. Rev. B, {\bf 87} 245408 (2013). arXiv:1211.4711
 
 
 
\bibitem{mattausch2007}A. Mattausch and O. Pankratov, Phys. Rev. Lett. {\bf 99}, 076802 (2007).

\bibitem{varchon2007}F. Varchon, et. al.
Phys. Rev. Lett. {\bf 99}, 126805 (2007)

\bibitem{pankratov} O. Pankratov, S. Hensel, and M. Bockstedte, Phys. Rev. B {\bf 82}, 121416(R) (2010)
 
\bibitem{giovannetti2007} G. Giovannetti, P. A. Khomyakov, G. Brocks, P. J. Kelly, and J. van den Brink, Phys. Rev. B {\bf 76}, 073103 (2007). 

\bibitem{zhou2007} S. Y. Zhou, et. al.,
Nat. Mat. {\bf 6}, 770 (2007).



\bibitem{kindermann2012} 
M. Kindermann, B. Uchoa and D. L. Miller , Phys. Rev. B {\bf 86}, 115415 (2012); arXiv:1205.3194 






\bibitem{guinea2010}
F. Guinea, T. Low,
Phil. Trans. R. Soc. A {\bf  368}, 5391 (2010).





\bibitem{Park2008_NatPhys} C. H. Park, L. Yang, Y. W. Son, M. L. Cohen, S. G. Louie, Nature Phys. {\bf 4}, 213 (2008) 

 \bibitem{Park2008_PRL} C. H. Park, L. Yang, Y. W. Son, M. L. Cohen, S. G. Louie, Phys. Rev. Lett. {\bf 101}, 126804 (2008).












\bibitem{amet2012} F. Amet, J. R. Williams, K. Watanabe, T. Taniguchi, D. Goldhaber-Gordon, 
Phys. Rev. Lett., {\bf 110} 216601 (2013).

\bibitem{hunt2013} B. Hunt, et. al., 
Science, {\bf 340} 1427-1430 (2013).

\bibitem{cory-aps2013} Cory Dean, 
APS March Meeting, M2.00003 (2013).


\bibitem{kotov2009}
V. N. Kotov, B. Uchoa, and A. H. Castro Neto, Phys. Rev. B 80, 165424 (2009)





\bibitem{sachs2011} B. Sachs, T. O. Wehling, M. I. Katsnelson, and A. I. Lichtenstein, Phys. Rev. B {\bf 84}, 195414 (2011). 

\bibitem{yankowitz2012} M. Yankowitz, et al., Nature Phys. {\bf 8}, 382 (2012).



\bibitem{gonzalez1994} J. Gonz\'alez, F. Guinea, and M. A. H. Vozmediano, Nucl. Phys. B {\bf 424}, 595 (1994); Phys. Rev. B {\bf 59}, R2474 (1999).


 \bibitem{vafek2007} O. Vafek, Phys. Rev. Lett. {\bf 98}, 216401 (2007).

\bibitem{son2007}
D. T. Son, Phys. Rev. B {\bf 75}, 235423 (2007).




\bibitem{elias2011} D. C. Elias, R. V. Gorbachev, A. S. Mayorov, S. V. Morozov, A. A. Zhukov, P. Blake, L. A. Ponomarenko, I. V. Grigorieva, K. S. Novoselov, F. Guinea, and A. K. Geim, 
Nature Phys. {\bf 7}, 701 (2011).



\bibitem{levitov2001} V. I. Talyanskii, D. S. Novikov, B. D. Simons, and L. S. Levitov, Phys. Rev. Lett. {\bf 87}, 276802 (2001)


\bibitem{levitov2003} L. S. Levitov and A. M. Tsvelik, Phys. Rev. Lett. {\bf 90}, 016401 (2003).





\bibitem{batterfly_manchester}
L. A. Ponomarenko,  et al.
Nat. Phys. {\bf 497} 594 (2013). arXiv: 1212.5012





\bibitem{dean2010} C. R. Dean,  et al.  Nature Nanotech. {\bf 5}, 722 (2010). 




 
 \bibitem{ponomarenko2011} L. A. Ponomarenko,  et al.
Nature Phys. {\bf 7}, 958-961(2011).





 
 










%
%
%
%
%







\end{thebibliography}
\end{document}